\documentclass[prl,twocolumn]{revtex4}
\usepackage{graphicx}
\renewcommand{\v}[1]{{\bf #1}}
\renewcommand{\b}[1]{\bar{#1}}
\newcommand{\sign}{{\rm sign}}

\def\eqa{\begin{eqnarray}}
\def\eea{\end{eqnarray}}
\newcommand{\eq}{\begin{equation}}
\newcommand{\ee}{\end{equation}}

\renewcommand{\>}{\rangle}

\renewcommand{\Im}{{\rm Im}}

\newcommand{\ua}{\uparrow}
\newcommand{\da}{\downarrow}
\newcommand{\ra}{\rightarrow}

\newcommand{\cK}{ {\cal K} }

\begin{document}

\title{Theory of doped Mott insulators: duality between pairing and magnetism}
\author{Qiang-Hua Wang}
\affiliation{National Laboratory of Solid State Microstructures,
Institute for Solid State Physics, Nanjing University, Nanjing
210093, China}


\begin{abstract}
By bosonizing the electronic t-J model exactly on any
two-dimensional (2D) lattices, and integrating out the gauge
fluctuations combined to slave particles beyond mean fields, we
get a theory in terms of physical Cooper pair and spin
condensates. In the sense of mutual Berry phase they turns out to
be dual to each other. The mutual-duality is the missing key in
the resonant-valance-bond idea\cite{rvb} to work as a paradigm of
doped 2D Mott insulators. We argue that essential aspects of
high-$T_c$ phenomenology find natural solutions in the theory. We
also provide interesting predictions for systems on hexagonal
lattices.
\end{abstract}

 \maketitle

{\bf\em I. Introduction:} Ever since Anderson's proposal that
superconductivity in high-$T_c$ copper oxides may be a realization
of resonant-valence-bond (RVB) physics,\cite{rvb} solving the t-J
model,\cite{theories,phasestring} a standard electron model for
systems with strong on-site Coulomb repulsion, has been not only a
daydream but also a nightmare in the community for more than a
decade. The good news is that indeed a lot of high-$T_c$
phenomenology may be interpreted in one way or another using the
t-J model. The bad news is a consensus to all phenomenology is not
reached even just in the underdoped region. It is therefore
academically desirable to have a theoretical paradigm (in the
infrared limit), just as Luttinger-liquid theory in
one-dimensional (1D) systems, Landau-Fermi liquid theory for
metals and quantum Hall liquid for semiconductor wells under high
magnetic fields. On the theory side such a paradigm helps to
answer the question: Does t-J model describe high-$T_c$ anyway? If
it does, what makes high-$T_c$ unique? Inspired by the fact that
Luttinger liquid theory can be easily brought out by bosonization
in 1D and recent successes of the phase-string
theory,\cite{phasestring} in this Letter we tackle the challenge
raised by Anderson to generalize the Luttinger liquid to 2D
strongly correlated electron systems.\cite{andersonbook} We
bosonize the electronic t-J model exactly [see Eqs.(\ref{hj}) and
(\ref{ht})]. The mean field theory integrates spin ordering and
pairing thanks to the ``all-boson'' hamiltonian. We further
integrate out gauge fluctuations beyond mean field and obtain a
theory in terms of physical Cooper pair and spin magnetic
condensates alone [see Eqs.(\ref{lcs}), (\ref{lc}), and
(\ref{lm})]. They turn out to be dual to each other in the sense
that Cooper pair condensate (spin condensate) sees the classical
part of magnetic moments (Cooper pairs) as $2\pi$-flux
bundles.\cite{dual}. The mutual-duality between superconductivity
and magnetism is clearly what makes doped 2D Mott insulators
unique, and is just the missing key in the RVB theory. Because of
the intervening magnetism, it seems unnecessary to prepare a
quantum spin liquid in parent Mott insulators to adiabatically
connect superconductivity. Other far-reaching consequences of the
theory are detailed in the closing section after warming up with
the notations and technical expositions, which could be skipped
for anxious readers.

{\bf\em II. Bosonization of the t-J model:} The electronic t-J
model in the so-called slave-boson representation can be written
as \eqa H &=& H_t
+H_J-\mu\sum_i (1-h_i^\dagger h_i),\\
H_t&=&-t\sum_{\<ij\>}K_{ij}^\dagger K^h_{ij}+{\rm h.c.},\\
H_J&=&-\frac{J}{2}\sum_{\<ij\>}K^\dagger_{ij}K_{ij}
=-\frac{J}{2}\sum_{\<ij\>}P^\dagger_{ij}P_{ij},\eea where \eqa
K^h_{ij}&=&h_i^\dagger h_j,\ K_{ij}=\sum_\sigma
f^\dagger_{i\sigma}f_{j\sigma},\ P_{ij}=\sum_{\sigma}\sigma
f_{i\sigma}f_{j\bar{\sigma}}.\eea In the above $h_i$
($f_{i\sigma}$) is hardcore holon (fermionic $\sigma$-spinon
annihilation operators). We have written $H_J$ in the hopping and
singlet pairing channels. {\it They help identify the saddle point
of the theory in any dimension and for any lattice geometry}. The
physical Hilbert space is a subspace of the Fock space subject to
\eq h^\dagger_i h_i+\sum_\sigma f_{i\sigma}^\dagger
f_{i\sigma}\equiv 1.\ee The slave-boson hamiltonian is invariant
under the local U(1) gauge transform \eqa f_{i\sigma}&\ra&
f_{i\sigma}\exp(i\theta_i),\ h_i \ra h_i \exp(i\theta_i).\eea This
symmetry guarantees the conservation of the local gauge charge
$h_i^\dagger h_i +\sum_\sigma f_{i\sigma}^\dagger f_{i\sigma}$ and
the associated current.

While not necessarily unique, we define the following
Jordan-Wigner transformation \eqa f^\dagger_{i\sigma}\ra N_\sigma
b^\dagger_{i\sigma}\exp\Big[i\sigma\sum_{l\neq i}\theta_{il}
(n_{l\sigma}-\b n_\sigma)\Big],\eea  were $N_\sigma$ is the Klein
factor, $b_{i\sigma}$ hardcore boson annihilation operator,
$\theta_{il}$ is a multi-valued function defined by \eqa
\theta_{il}=\Im\ln (z_i-z_l),\ z_i=x_i+iy_i,\eea
$n_{l\sigma}=b^\dagger_{l\sigma}b_{l\sigma}$ and finally $\b
n_\sigma$ is the average occupation of $\sigma$-spinons. For
systems with anti-ferromagnetic exchange it is reasonable to set
$\b n_\sigma=(1-x)/2$ with $x$ the holon concentration. The
purpose of including $\b n_\sigma$ is to make the gauge field
attached to the $b_\sigma$-bosons vanish at the mean field level.

By simple substitution and taking into account of the hardcore
nature of the bosons as well as the occupation constraint, $H_J$
can be rewritten as \eqa
H_{J}&=&-\frac{J}{2}\sum_{\<ij\>}\cK_{ij}^\dagger\cK_{ij}, \label{hj}\\
\cK_{ij}&=&\sum_{\sigma}b^\dagger_{i\sigma}b_{j\sigma}
\exp[-i\sigma(a^h_{ij}+\eta_{ij})],\\
a^h_{ij}&=&\frac{1}{2}\sum_{l\neq
ij}(\theta_{il}-\theta_{jl})(n^h_l-x),\\
\eta_{ij}&=&\frac{1-x}{2}(\theta_{ij}-\theta_{ji}). \eea Here
$n^h_l=h^\dagger_lh_l$. The bond phase $\eta_{i,i\pm \v
\delta}=\mp (1-x)\pi/2$ for $\v \delta=\hat{x}$ and $ \hat{y}$ on
square lattices, and for $\v \delta=\hat{x}$ and $
\sqrt{3}\hat{y}/2\pm \hat{x}/2$ on hexagonal lattices. From the
above we see that {\em holons appear as $\pi$-flux bundles as
viewed by spinnons}. This effect is cancelled by the background
field due to $x$ at the mean field level, but should be kept
beyond the mean field level due to its topological nature.
Certainly the above definition of ${\cal K}_{ij}$ is unique up to
arbitrary spin-independent link phase factors, \eqa {\cal
K}_{ij}&\ra& {\cal K}_{ij}\exp(-ia_{ij}),\eea for any $a_{ij}$, a
reflection of the local U(1) gauge symmetry. We choose not to
present the pseudo-pairing operator at this stage. As will be
clear, {\em pairing is an inevitable mode in the theory}.

Now we can proceed to transform the kinetic part $H_t$. By
substitution and manipulating in terms of ${\cal K}_{ij}$, we get
\eqa H_t&=&-t\sum_{\<ij\>}{\cal K}_{ij}^\dagger {\cal
K}^h_{ij}+{\rm
h.c.},\label{ht} \\ {\cal K}^h_{ij}&=&h_i^\dagger h_j \exp(-ia^m_{ij}),\\
a^m_{ij}&=&\frac{1}{2}\sum_{l\neq
ij}(\theta_{il}-\theta_{jl})(n_{l\ua}-n_{l\da}), \eea where we
have used the following identity \eq
\sigma\Big[n_{l\sigma}+\frac{1}{2}(n_{lh}-1)\Big]\equiv
\frac{1}{2}(n_{l\ua}-n_{l\da}).\ee {\em Therefore z-direction
magnetic moments appear as $\pi$-flux bundles as viewed by
holons}. Again the above definition of ${\cal K}^h$ is unique up
to the same U(1) gauge transform applied to ${\cal K}_{ij}$.

Summarizing, $H=H_t+H_J-\mu\sum_i(1-n^h_i)$ with the new forms of
$H_t$ and $H_J$ in Eqs.(\ref{hj}) and (\ref{ht}) discussed above
forms another rigorous fully-bosonized representation of the t-J
model. The holons and spinnons are mutually frustrated as
reflected by the gauge fields $a^h$ and $a^m$. We note that a
similar notion was pointed out in the phase string theory designed
particularly for bipartite lattices.\cite{phasestring} The present
theory applies for any lattice geometries.

{\bf\em III. Mean field theory:} The interacting part of the
hamiltonian can be rewritten as $
H_t+H_J=-\frac{J}{2}\sum_{\<ij\>}H^\dagger_{ij}H_{ij}
+\frac{2t^2}{J}\sum_{\<ij\>}n^h_in^h_j$ with $
H_{ij}=\cK_{ij}+\frac{2t}{J}\cK^h_{ij}$. This suggests the MF
effective hamiltonians for holons and spinons, respectively, \eqa
H^h_{MF}&=&-t\sum_{\<ij\>}(\chi^*_{ij}\cK^h_{ij}+{\rm
h.c.})+\mu\sum_i n^h_i,\\
H^s_{MF}&=&-\frac{J}{2}\sum_{\<ij\>}(\chi^*_{ij}\cK_{ij}+{\rm
h.c.})-\lambda\sum_{i\sigma}n_{i\sigma},\eea where we suppressed
the residual interaction for brevity. Here $\chi_{ij}$ is the mean
field hopping amplitude $\chi_{ij}=\<H_{ij}\>$. The Lagrangian
multipliers $\mu$ and $\lambda$ enforces $\<h^\dagger_ih_i\>=x$
and $\sum_\sigma \<b^\dagger_{i\sigma} b_{i\sigma}\>=1-x$,
respectively. Let us assume zero average magnetization in the
z-direction so that both holons and spinons see zero flux in the
MF theory. An obvious mean field solution is $\chi_{ij}=s|\chi|$
with $s=\sign(t)$ so that all three species of bosons can condense
to their own {\it non-degenerate} ground states: $h_i=\sqrt{x}$,
and $b_{i\sigma}=\sqrt{\frac{1-x}{2}}\exp[-i\sigma s\frac{\v
Q}{2}\cdot\v r_i]$ with $\v Q=\v Q_0(1-sx)$. Here $\v
Q_0=(\pi,\pi)$ and $2\pi(1/3,1/\sqrt{3})$ on a square and
hexagonal lattice, respectively. By self consistency,
$|\chi|=1-x+2|t|x/J$ and $(1-x)\sin[(1-sx/2)\pi/3]+2|t|x/J$ on
square and hexagonal lattices, respectively. $\v Q$ would be the
magnetic ordering wave vector as seen from
$\<S^+_i\>=\<b^\dagger_{i\ua}b_{i\da}\>=\frac{1}{2}(1-x)\exp(is\v
Q\cdot \v r_i)$. The $s$-dependence in $\v Q$ on square lattices
is trivial because $\v Q_0(1\pm x)$ are identical, but is
nontrivial on hexagonal lattices. One would also thought of spin
pairing $b_{i\ua} b_{j\da}$ in the condensates. The bosonized MF
state therefore integrates hopping, magnetism and spin pairing.
However, we must remind ourselves that the mean field breaks the
local U(1) gauge symmetry. It is a must to integrate out gauge
fluctuations beyond mean field theory to talk about physical
entities. This is the subject of the gauge theory in the next
section.

{\bf\em IV. Integration over gauge fluctuations:} The Euclidian
action $S=\int L d^2xd\tau$ that describes the long-wave-length
low-energy fluctuations beyond the above MF state is as follows.
The Lagrangian density $L=L_h+\sum_\sigma L_\sigma+L_{CS}+ia_0$
with $L_h$, $L_\sigma$ and $L_{CS}$ describe, respectively,
holons, $\sigma$-spinons and the Chern-Simons-like flux-attaching
reflecting the mutual frustration between holons and spinons. The
holon part is given by \eqa L_h&= &i(\b \rho_h+\delta
\rho_h)(\phi_h^*\frac{\partial_\tau}{i}\phi_h-a_0+A_0)
-i\delta\rho_h a^h_0+\frac{u_h}{2}\delta\rho_h^2\nonumber\\&+&i\v
j_h\cdot(\phi^*_h\frac{\nabla}{i}\phi_h-\v a-\v a^m+\v A)+
\frac{1}{2K_h}\v j_h^2.\eea In the above $\b\rho_h=x$ is the
average holon concentration, $\delta\rho_h$ is the holon density
fluctuation, $\v j_h$ is the holon current density, $K_h\propto
2tx\chi$ is the bare holon phase stiffness (this is best seen by
integrating out $\v j_h$), $\phi_h$ is defined by $h=|h|\phi_h$,
$(a_0,\v a)$ is the auxiliary U(1) gauge field, $a^h_0$ is a
Lagrangian multiplier for flux attaching to frustrate spinons, $\v
a^m$ is the gauge field due to the frustration from spinons,
$(A_0,\v A)$ is the physical electro-magnetic (EM) gauge field
coupled to the positively-charged holons, and finally $u_h$ is an
effective short-range repulsion between holons.  Similarly, the
$\sigma$-spinon part $L_\sigma$ is given by \eqa L_\sigma&=&i(\b
\rho_\sigma+\delta
\rho_\sigma)(\phi_\sigma^*\frac{\partial_\tau}{i}\phi_\sigma-a_0)
-i\sigma\delta\rho_\sigma a^m_0
+\frac{u_b}{2}\delta\rho_\sigma^2\nonumber\\&+&i\v
j_\sigma\cdot(\phi^*_\sigma\frac{\nabla}{i}\phi_\sigma-\v
a-\sigma\v a^h)+\frac{1}{2K_b}\v j_\sigma^2 .\eea In the above
$a^m_0$ is a Lagrangian multiplier that attach flux to the
z-direction magnetization that frustrates holons, $K_b\propto
J\chi (1-x)/2$ is the bare spinon phase stiffness, and
$\phi_\sigma$ is defined by $b_\sigma=|b_\sigma|\exp(-i\sigma\v
Q\cdot r/2)\phi_\sigma$ which describes fluctuations of the
$\sigma$-boson beyond the condensate momentum. The last piece
$L_{CS}$ is given by \eqa L_{CS}&=&
\frac{i}{\pi}a^h_0(\nabla\times\v
a^h)_z+\frac{i}{\pi}a^m_0(\nabla\times\v a^m)_z.\label{lcs}\eea In
the above the subscript $z$ means the $z$-component of the object.
Let us observe that integration over $a_0$, $a^h_0$ and $a^m_0$
using the total action $L$ indeed enforces occupation constraint
and attaches $\pi$-flux to holon and magnetization fluctuations,
respectively.

Given the above form of the action, we can integrate out the
auxiliary U(1) gauge field $(a_0,\v a)$ exactly
,\cite{dhleeqhwang} yielding
$\delta\rho_h+\sum_\sigma\delta\rho_\sigma=0$ and $\v
j_h+\sum_\sigma\v j_\sigma=0$. This enables us to get rid of slave
particles in favor of a U(1) gauge-neutral theory
$L=L_c+L_m+L_{CS}$ as follows.

First, $L_c$ describes the charge condensate (CC) of the theory,
\widetext \eqa & &L_c=i \b
\rho_c(\phi_c^*\frac{\partial_\tau}{i}\phi_c-2A_0)+iA_0
+i\delta\rho_c(\phi_c^*\frac{\partial_\tau}{i}\phi_c+2a^h_0-2A_0)
+i\v j_c\cdot(\phi_c^*\frac{\nabla}{i}\phi_c+2\v a^m-2\v
A)+\frac{u_c}{2}\delta\rho_c^2+\frac{1}{2K_c}\v
j_c^2,\label{lc}\\& & \b\rho_c=\frac{1}{2}(1-x),\
\phi_c=\phi_\ua\phi_\da(\phi_h^*)^2, \
\delta\rho_c=\frac{1}{2}\sum_\sigma\delta\rho_\sigma,\ \v
j_c=\frac{1}{2}\sum_\sigma\v j_\sigma,\
\frac{1}{K_c}=\frac{2}{K_b}+\frac{4}{K_h}, \
u_c=2u_b+4u_h.\label{kc}\eea Here we have used the identity
$\sum_\sigma\phi^*_\sigma\partial_t\phi_\sigma-2\phi^*_h\partial_t\phi_h\equiv
\phi_c^*\partial_t\phi_c$. Clearly $\phi_c$ is the phase factor of
$b_\ua b_\da h^* h^*$, a spin-charge recombined Cooper pair.
Moreover, $\b\rho_c$ and $\v j_c$ are just the Cooper pair density
and spatial current, respectively, $K_c$ is the effective
zero-temperature superfluid density, and finally $u_c$ is the
effective local charge repulsion. At low doping $K_c\propto x$.
Interestingly, $A_0$ in the first two terms of $L_c$ simply
reminds us that {\em the effective charge density is $1-2\b
\rho_c=x$, but the Cooper pairs response to EM field as
$2e$-carriers with pair density $\b \rho_c=(1-x)/2$}. This proves
that pairing does not imply bounding of holes at all (they are not
the normal modes in the theory), a conclusion supported by
variational Monte Carlo studies of the t-J+Coulomb
model.\cite{mottpair} Moreover,{\em the Cooper pairs also view a
localized z-direction magnetic moment as a $2\pi$-flux bundle as
described by $2\v a^m$.}

Second, $L_m$ describes the spin magnetic condensate (MC) of the
theory, \eqa & &L_m=i
\delta\rho_m(\phi_m^*\frac{\partial_\tau}{i}\phi_m-2a^m_0) +\v
j_m\cdot(\phi_m^*\frac{\nabla}{i}\phi_m-2\v
a^h)+\frac{u_m}{2}\delta\rho_m^2+\frac{1}{2K_m}
\v j_m^2,\label{lm}\\
& &\delta\rho_m=\frac{1}{2}\sum_\sigma\sigma\delta\rho_\sigma, \
\phi_m=\phi_\ua\phi_\da^*, \ \v j_m=\frac{1}{2}\sum_\sigma\sigma\v
j_\sigma, \ K_m=\frac{1}{2}K_b,\ u_m=2u_b.\eea
\endwidetext
Clearly, $\delta\rho_m$, $\phi_m$ and $\v j_m$ are the
z-direction magnetization $S^z$, the phase factor of $S^-$ and the
spin current (beyond the magnetic ordering wave vector $\v Q$).
The gauge field $2\v a^h$ describes the effect of Cooper pair
density fluctuations on the spin dynamics. A localized Cooper pair
would appear as a $2\pi$-flux bundle to $S^-$. On the other hand,
MC condensate is not coupled to physical EM gauge fields and is
therefore charge-neutral, as would be desirable.

Except for the first two terms in $L_c$ (which gives the average
density of Cooper pairs and the charge carrier density) the two
condensates CC and MC are {\it mutually dual} to each other. This
is the crucial part of the theory that makes a difference to the
1D counterpart in which spin-charge separation is complete.

{\bf\em V. Discussions:} The following discussion at zero
temperature relies on the notion of mutual-duality identified
above and the self-duality for a separate condensate
alone.\cite{dual,duals} We specialize to the context of high-$T_c$
phenomenology, \cite{hightc} but the basic physics applies to all
doped Mott insulators. 1) At zero doping we have a quantum
antiferromagnetism described by the MC alone. The gauge field $\v
a^h=0$ since there is no charge fluctuations at all. 2) At low
doping CC is disordered by quantum (or thermal) fluctuations due
to the small phase stiffness. The charges are localized and act as
vortices to MC by mutual-duality. Depending on whether MC is of
type-I or type-II ``superconductor'', these charges may either
phase separate or form a Wigner crystal, respectively. They
correspond to the intermediate state and mixed state in usual
superconductors, and would suggest two types of Mott insulators as
recently advocated.\cite{motttypes} This is a useful concept to
understand the absence or presence of electronic inhomogeneities
in different families of high-$T_c$ copper
oxides.\cite{inhomogeneity} Localized charges can make the spin
order glassy. Another interesting possibility is the charge stripe
order in cuprates. Since the localized charges appear as vortices
to MC, the fact that stripe is an anti-phase boundary to the
underlying magnetic order \cite{antiphase} seems very natural. 3)
With increasing doping, the charges are more and more mobile and
eventually CC shows up above some lower critical doping level. 3a)
As we envisioned above, although the doped holes do not pair at
all, the Cooper pairs response to EM fields as charge-2e carriers.
This determines the correct flux quantum as $hc/2e$. 3b) According
to the composition rule of superfluid density $K_c$ in
Eqs.(\ref{kc}), for small $x$ it is dominated by the smaller
component $K_h$, and is therefore roughly linear in $x$. 4)
Consider ordered CC and disordered MC, a superconductor with no
spin order. The vortices of MC must condense by
self-duality.\cite{dual} Let us call this new condensate as
M$^\prime$C. Just as CC does, M$^\prime$C also views localized
magnetization $S^z$ as $2\pi$-flux bundles, {\em i.e.}, the next
generation vortex excitations. Therefore CC and M$^\prime$C can
happily coexist. This opens up a handful of interesting
possibilities. 4a) Vortices under an applied magnetic field in CC
may capture local spin magnetic moments, forming non-topological
vortices with zero winding number in CC to lower the superflow
kinetic energy. Since winding is a necessary condition to the zero
energy electron state in a vortex core of d-wave
superconductor,\cite{jh} non-topological vortices may resolve the
puzzle that no vortex-core zero-bias tunnelling conductance peak
has ever been observed in cuprates.\cite{corestm} 4b) A vortex in
CC may also capture many spin moments (favorably ordered with
staggered polarization), and this would lead to the checkerboard
pattern in the local density of states near a vortex core in
cuprates.\cite{hoffman} 4c) Moreover, since the captured spins are
vortices to M$^\prime$C by self-duality, it is favorable to
recover MC. This may explain the enhancement of anti-ferromagnetic
correlation in the vortex state of cuprates.\cite{lake} 4d) The
spin excitations $(\delta\rho_m,\v j_m)$ appear as the ``EM''
fields to M$^\prime$C by self-duality.\cite{dual} Such ``photons''
are gapped up to the plasma frequency of CC+M$^\prime$C state that
scales with the physical superfluid density. In view of our
starting magnetic vector $\v Q$, this gap actually corresponds to
the inelastic neutron resonance energy found in cuprates.
\cite{resonance} The doping dependent incommensurability in $\v Q$
is obvious in the theory. 4e) On the other hand, assuming the
absence of long-range interactions in either CC and/or
M$^\prime$C, longitudinal phase fluctuations in the condensates
are phonon-like, and via re-fermionization of the emerging low
energy field theory\cite{pending} they are nothing but the
Dirac-like nodal quasi-particles in a d-wave superconductor. 4f)
Topologically de-confined but spatially localized spin moments,
such as caused by Zn, induces vortices in CC at zero magnetic
field by mutual-duality. Therefore the critical Zn concentration
to kill CC would be given by $H_{c2}/\Phi_0$. (The spins do not
have to point in the same direction.) This seems to agree with the
so-called ``Swiss cheese'' model of Zn in cuprates.\cite{zinc}
More detailed discussions are in progress.

\acknowledgements{I thank Z. Y. Weng for enlightening discussions
and encouragements. This work is supported by NSFC 10204011 and
10021001.}


\begin{references}
\bibitem{duals} Strictly speaking self-duality is defined only at $T=0$, but mutual-duality applies even at finite temperatures.
\end{references}
\end{document}